\documentclass[12pt]{article} 
\usepackage{epsf}                                       
\newcommand{\newc}{\newcommand}    
\newc{\ra}{\rightarrow} 
\newc{\lra}{\leftrightarrow} 
\newc{\beq}{\begin{equation}} 
\newc{\eeq}{\end{equation}} 
\newc{\barr}{\begin{eqnarray}} 
\newc{\earr}{\end{eqnarray}} 
\newc{\texa}{\textstyle}
\newc{\paral}{\parallel}
\newc{\und}{\underline}
\newc{\pars}{\partial}
\newc{\nonu}{\nonumber \\}

\begin{document}
\thispagestyle{empty}
\begin{center}
{\Large \bf 
On resistive magnetohydrodynamic  equilibria  \\ \vspace{3mm}  
of an axisymmetric toroidal plasma  with flow \vspace{4mm} }\\ 
\large\bf
{\large G. N. Throumoulopoulos  \footnote             
{Permanent  address: 
University of Ioannina, Association 
EURATOM / Hellenic Republic, Physics Department,
Section of Theoretical Physics, 
GR  451 10 Ioannina, Greece} and H. Tasso  \vspace{1mm}\\
{\it   Max-Planck-Institut f\"{u}r Plasmaphysik, EURATOM
Association \\ \vspace{1mm}
 D-85748 Garching, Germany }\\ \vspace{3mm}
October 1999}  
\end{center}
%
%
\begin{center}
{\large \bf Abstract}
\end{center}

It is shown that the magnetohydrodynamic  equilibrium
states   of an axisymmetric toroidal plasma with finite resistivity 
and flows parallel to the magnetic field  are governed by
a second-order partial differential equation for the poloidal magnetic
flux function $\psi$ coupled with a  Bernoulli type equation for the plasma 
density  
(which are identical in form to the corresponding ideal MHD
equilibrium equations) along with the relation
$\Delta^\star \psi=V_c \sigma$.  (Here, $\Delta^\star$ is the 
Grad-Schl\"{u}ter-Shafranov operator, $\sigma$
is the conductivity and $V_c$ is the constant toroidal-loop voltage   
divided by $2 \pi $).
In particular, for incompressible flows the
above mentioned partial differential equation becomes elliptic and decouples
 from
the Bernoulli equation   
[H. Tasso and G. N. Throumoulopoulos, Phys. Plasmas {\bf 5}, 2378 (1998)].
For a conductivity of the form  $\sigma=\sigma(R, \psi)$
($R$ is the distance from the axis of symmetry)
 several classes of analytic equilibria with incompressible flows 
can be constructed having qualitatively plausible $\sigma$ profiles,
i.e. profiles with  
 $\sigma$ taking a  maximum value  close to the magnetic axis and a minimum value on
the plasma surface. For $\sigma=\sigma(\psi)$  consideration of the relation 
$\Delta^\star\psi = V_c \sigma(\psi)$
in the vicinity of the magnetic axis leads therein to a proof of 
the non-existence 
of either compressible or incompressible equilibria. This result can be extended
to the more general case  of non-parallel flows lying within the 
magnetic surfaces.
\vspace{1mm}


\newpage

\begin{center} 
{\large \bf I.\ \ Introduction}
\end{center}
 
In addition to the case of the long living astrophysical plasmas, 
understanding   the equilibrium  properties of resistive
fusion  plasmas is  important, particularly in view of  
the next step devices which will possibly demand pulse 
lengths of the order of $10^3$ secs
(or more for an ITER size machine)  (\cite{MoVo} and Refs. cited therein).  
Theoretically, however, it was proved by Tasso \cite{Ta} 
that  resistive equilibria with   $\sigma=\sigma(\psi)$
 are not compatible with 
the Grad-Schl\"uter-Shafranov equation. (Here, 
$\sigma$ is the conductivity and $\psi$ is the poloidal magnetic
flux function.)
The non-existence
of static axisymmetric resistive equilibria with a uniform conductivity was
also  suggested  recently \cite{MoSh,BaLe,MoBh}. 
Also, in the 
collisional regime Pfirsch and Schl\"uter  
showed that the toroidal curvature gives  rise to 
an enhanced diffusion, 
which is related to the  conductivity parallel to the magnetic field. 
In the above mentioned studies
the inertial-force flow term 
$\rho({\bf v}\cdot\nabla){\bf v}$  is neglected in the equation of 
momentum conservation.
For ion flow velocities of the order of $100$ Km/m,
 which have
been observed
in neutral-beam-heating experiments \cite{SuEu,BrBi,TaDo}
the term
$\rho({\bf v}\cdot\nabla){\bf v}$ 
can not be considered negligible. Therefore, it is
worthwhile to investigate the nonlinear 
resistive equilibrium, in particular to address    
the following issues: (a) the impact of 
the non-linear flow in the Pfirsch-Schl\"uter diffusion, and 
 (b) the existence of resistive equilibria, in particular 
equilibria with $\sigma=\sigma(\psi)$.
Since the  magnetohydrodynamic (MHD) equilibrium with arbitrary  flows 
and finite conductivity is a very difficult problem,
in  a recent study \cite{Th} we considered  an 
axisymmetric toroidal plasma 
with purely toroidal flow including the term $\rho({\bf v}\cdot \nabla)
{\bf v}$ in the momentum-conservation equation. It was shown that  
the non-linear flow does not affect the static-equilibrium 
situation, i.e  $\sigma=\sigma(\psi)$ equilibria are not
possible.

A  way of constructing more plausible  equilibria 
from the physical point of view 
could be by  considering  flows  less restricted in direction. 
Taking also into account the fact 
that the poloidal flow in the edge region of magnetic-confinement  systems
plays a role in the transition from the low-confinement mode to
the high-confinement mode, in the present report we extend  
our previous studies  
to the case of flows having  non-vanishing poloidal components in addition
to toroidal ones. 
Because of the difficulty of the problem
we consider flows parallel to the magnetic field.
Some of the conclusions, however, can be extended to non-parallel
flows lying within the magnetic surfaces.
It is also noted that  possible equilibria with parallel flows 
would be free of Pfirsch-Schl\"uter diffusion because 
the convective term ${\bf v}\times {\bf B}$ in the Ohm's low
vanishes.  The main conclusion is that  
for the system under consideration the existence of  
equilibria depends crucially on 
the  spatial dependence of conductivity. The report
is organized as follows. The equilibrium equations  
for
an axisymmetric toroidal resistive plasma with 
parallel flows surrounded by a conductor
are derived in Sec. II. The existence of solutions  
is then examined in Sec. III for the cases $\sigma=\sigma(R, \psi)$
($R$ is the distance from the axis of symmetry), and 
$\sigma=\sigma(\psi)$.  Sec. IV summarizes our conclusions.

\begin{center}
{\large\bf II.\ \ Equilibrium equations}
\end{center}

The MHD equilibrium states of a plasma
with scalar conductivity
are governed by the following set of
equations, written in standard notations and convenient units:
\begin{equation}
{\bf\nabla} \cdot (\rho {\bf v}) = 0,
					    \label{1}
\end{equation}
\begin{equation}
\rho ({\bf v} \cdot {\bf\nabla})  {\bf v} = {\bf j}
\times {\bf B} - {\bf\nabla} P,
					    \label{2}
\end{equation}
\begin{equation}
{\bf\nabla} \times  {\bf E} = 0,
					    \label{3}
\end{equation}
\begin{equation}
{\bf\nabla}\times {\bf B} = {\bf j },
					    \label{4}
\end{equation}
\begin{equation}
{\bf\nabla} \cdot {\bf B} = 0,
					    \label{5}
\end{equation}
\begin{equation}
{\bf E} +{\bf v} \times {\bf B} = \frac{\bf j}{\sigma}.
					    \label{6}
\end{equation}
It is pointed out that, unlike to the usual procedure  
followed in equilibrium studies  with flow
\cite{ZeGr,MoSo,Ha,SeGr,KeTo,ZeSt}  
in the present work
an equation of state is not included
in the above set of  equations from the outset  and therefore 
the equation of state independent Eqs. (\ref{13}) and (\ref{14})
below are first derived.  This alternative procedure is convenient
because the equilibrium problem is then further reduced  
for specific   cases associated  with several  equations of state.

The system under consideration 
is a toroidal axisymmetric magnetically confined   plasma, 
which is surrounded by a conductor (see  Fig. 1 of Ref. \cite{Th}). 
With the use of
cylindrical coordinates
$R, \phi, z$  the position
of the surface of the conductor   is specified by some 
boundary curve in the ($R$, $z$) plane. 
The equilibrium quantities do not depend on the azimuthal 
coordinate $\phi$.  Consequently, the divergence free 
magnetic field ${\bf B}$ and current density ${\bf j}$
 can be expressed, with the aid of Ampere's low (\ref{4}),
in terms of the stream functions $\psi(R,z)$ and $I(R,z)$
as 
\begin{equation} 
{\bf B} = I \nabla \phi +
 \nabla\phi \times \nabla\psi,
					     \label{7}
 \end{equation}
 and
\begin{equation}
{\bf j} = \Delta^\star\psi \nabla\phi - \nabla\phi \times \nabla I.
					     \label{8}
\end{equation}
Here, 
$\Delta^\star$ is the elliptic operator defined by
 $\Delta^\star = R^2\nabla\cdot(\nabla /R^2)$
and
constant $\psi$ 
surfaces are  magnetic surfaces. Also, it is assumed that the plasma
elements flow solely along {\bf B}:
\beq
\rho {\bf v} = K {\bf B},                       \label{9}
\eeq
where $K$ is a function of $R$ and $z$.
Acting the divergence operator on Eq. (\ref{9}) 
and taking into   account  Eq. (\ref{1}) one obtains 
$\nabla K \cdot {\bf B} =0$.
						      \label{9a}
Therefore, the function $K$ is  a surface quantity: 
\beq
K=K(\psi).
							\label{9b}
\eeq
Another surface
quantity is identified from the toroidal component of the 
momentum conservation equation (\ref{2}):
\beq
\left(1-\frac{K^2}{\rho}\right) I = X(\psi).             \label{9c}
\eeq
 From Eq. (\ref{9c}) it follows that, unlike the case in static equilibria,
I is not (in general)  a surface quantity.
Furthermore, expressing the time independent electric field by
\beq
{\bf E} = -\nabla \Phi + V_c \nabla \phi,
						 \label{10}
\eeq
where $V_c$ is the constant toroidal-loop voltage divided by $2 \pi $,
the poloidal and toroidal components of Ohm's law (\ref{6}), respectively, 
yield
\beq
\nabla \Phi =\frac{\nabla \phi\times \nabla I}{\sigma}
							\label{11}
\eeq
and
\beq
\Delta^\star\psi = V_c \sigma= E_\phi R\sigma.
							\label{12}
\eeq
Here,  $E_\phi$ is the toroidal component of $\bf E$.
Eq. (\ref{12}) has an impact on the boundary conditions,
i.e. the component of ${\bf E}$
tangential to the plasma-conductor interface does not vanish. 
Therefore, the container can not be considered  perfectly
conducting.
Accordingly,  Ohm's law with  finite conductivity applied in the
vicinity of the plasma-conductor interface does not permit
the existence of a surface layer of current \cite{Ja}.
It is now assumed that the position of the conductor is such that
its surface 
coincides with the outermost of the closed magnetic surfaces. 
Thus, the condition
${\bf B}\cdot {\bf n}=0$, where {\bf n} is the outward unit vector
normal to the plasma surface,  holds in  the plasma-conductor
interface  and therefore the pressure P must vanish on the boundary.  
It is noticed
that this  is possible  only in equilibrium, 
because in the framework of resistive MHD time dependent
equations, the magnetic flux is not conserved. 
%
With the aid of equations  (\ref{7})-(\ref{9c}) 
the components of Eq. (\ref{2}) along ${\bf B}$ 
and perpendicular to a magnetic surface are put in the respective forms
\begin{equation}
{\bf B}\cdot \left\lbrack\nabla\left(\frac{K^2 B^2}{2\rho^2}\right) 
       +\frac{\nabla P}{\rho}\right\rbrack  =0
					      \label{13}
\end{equation}
and
\begin{eqnarray}
\left\{\nabla\cdot \left\lbrack\left(1 - \frac{K^2}{\rho}\right)
      \frac{\nabla \psi}{R^2} \right\rbrack
      + \frac{K}{\rho}\frac{\nabla K \cdot \nabla \psi}{R^2}
      \right\}|\nabla\psi|^2
      & & \nonu
      +
      \left\{\rho\nabla\left(\frac{K^2 B^2}{2\rho^2}\right) 
      + \frac{\nabla I^2}{2R^2}  - \frac{\rho}{2R^2}
	\nabla\left(\frac{IK}{\rho}\right)^2 + \nabla P \right\}
	\cdot \nabla \psi = 0. & &
					      \label{14}
\end{eqnarray}
 Eq. (\ref{14}) has  a singularity when  
\begin{equation}
\frac{K^2}{\rho}=1. 
						  \label{24}
\end{equation}
On the basis of Eq. (\ref{9}) for $\rho {\bf v}$ and the
definitions $v_{Ap}^2\equiv\frac{\textstyle |\nabla \psi|^2}
{\textstyle \rho}$ for the Alfv\'en velocity associated with the
poloidal magnetic field and the Mach number
\begin{equation}
 M^2\equiv\frac{v_p^2}{v_{Ap}^2}= \frac{K^2}{\rho},
						   \label{25} 
\end{equation}
Eq. (\ref{24}) can be written as
$
M^2= 1.
$
 
 Summarizing, the resistive MHD equilibrium of an axisymmetric
toroidal plasma with parallel flow is governed by the
set of Eqs. (\ref{12}), (\ref{13}) and  ({\ref{14}).  Owing to
the direction of the flow parallel to ${\bf B}$, Eqs. (\ref{13})
and (\ref{14}){\em do not contain the conductivity} and are
identical in form to the corresponding equations governing ideal
equilibria.  
Therefore, on the one hand,
several properties  of the ideal equilibria, e.g. 
the Shafranov shift of the magnetic surfaces and the detachment
of the isobaric surfaces from the magnetic surfaces (see the 
discussion following Eq. (\ref{23}) in Sec IIC) remain valid.
On the other hand, as will be shown in Sec. III,  
the conductivity $\sigma$ in Eq. (\ref{12}) plays an important role
on the existence of equilibria.

To reduce further  
equations (\ref{13}) and (\ref{14}),  
the starting set
of equations   (\ref{1})-(\ref{6})
 must be supplemented by 
an  equation of state, e.g. $P=P(\rho, T)$,  
along with
an equation determining the transport
of internal energy. 
Such a rigorous treatment, however,  makes the equilibrium problem 
very cumbersome. Alternatively, one can assume additional properties
for the magnetic surfaces  associated with either isentropic processes,
or isothermal processes, or incompressible flows. These three cases
are separately examined 
in the remainder of this section. 

\begin{center}
{\large\em A.\ \ Isentropic magnetic surfaces}
\end{center}

We consider a plasma with large but  
finite conductivity such that for  times  short compared with
the diffusion time scale, the dissipative   term
$\approx j^2/\sigma$ can be neglected.  
This permits one to  assume  conservation of the entropy: 
$\bf v\cdot \nabla S=0$, which on account of  Eq. (\ref{9}) 
leads to $S=S(\psi)$ 
($S$ is the specific entropy).
It is noted that  the case $S=S(\psi)$ was considered in 
investigations on ideal equilibria 
 with arbitrary flows \cite{MoSo,Ha} and purely toroidal 
 flows \cite{MaPe,ThPa}, as well as on resistive  
  equilibria with   purely toroidal flows \cite{Th}.
In addition, the plasma is assumed  
to being a perfect gas whose
internal energy density $W$ is simply proportional to the temperature.
Then, the  equations  for the  thermodynamic potentials lead to  
\cite{MaPe}
\begin{equation}
P = A(S)\rho^\gamma
						      \label{16}
\end{equation}
and
\begin{equation}
W =\frac{A(S)}{\gamma -1}\rho^{\gamma -1} = \frac{H}{\gamma}.
						      \label{17}
\end{equation}
Here, $A=A(S)$ is  an arbitrary function of  
$S$,
$ H=W + P/\rho$ is the specific enthalpy 
and $\gamma$ is the ratio of specific heats.
For simplicity and without loss of generality we choose the function $A$
to be identical with $S$. Consequently,
integration of Eq. (\ref{13})  yields
\begin{equation}
\frac{K^2B^2}{2\rho^2} + \frac{\gamma}{\gamma-1}S\rho^{\gamma-1}=H(\psi).
						 \label{18}
\end{equation}
Eq. (\ref{14}) reduces then to
\begin{equation}
\nabla\cdot \left\lbrack\left(1 - \frac{K^2}{\rho}\right)
      \frac{\nabla \psi}{R^2} \right\rbrack
 + ({\bf v}\cdot {\bf B}) K^\prime + \frac{B_\phi}{R} X^\prime
 + \rho H^\prime -\rho^\gamma S^\prime =0,  
						      \label{19}
\end{equation}
where the prime denotes differentiation with respect to $\psi$.
Apart from a factor $1/(\gamma -1)$ in the last term of the right-hand
side ($[1/(\gamma-1)]\rho^\gamma S^\prime$ instead of 
$\rho^\gamma S^\prime$)
Eq. (\ref{19}) is identical in form
with the  corresponding ideal MHD equation obtained by 
Hameiri \cite{Ha} (Eq. (7) therein).  
It should be noted that Eq. (22) remains regular 
for the case of isothermal plasmas ($\gamma=1$) while Hameiri's
result would make the equilibrium equation strangely   singular.           
In particular,
for $S=S(\psi)$ and $T=$ const. Eq. (\ref{16}) leads to $\rho=\rho(\psi)$ and
consequently the incompressibility equation $\nabla\cdot {\bf v}=0$ 
follows from  Eq. (\ref{1}). 
Incompressible flows, however,  are described by Eq. (\ref{26})
below which is free of the above mentioned singularity.
 
Unlike the case of static equilibria, Eq. (\ref{19}) is not
always elliptic; there are three critical values of  
the poloidal-flow Mach-number $M^2$
at which the type of this equation changes, i.e. it becomes
alternatively elliptic and hyperbolic \cite{ZeGr,Ha}. The toroidal flow
is not involved in these  transitions  because this is incompressible
by axisymmetry and, therefore,  does not relate to hyperbolicity
(see also the discussion in the beginning of Sec. IIC).

\begin{center}
{\large\em B.\ \ Isothermal magnetic surfaces}
\end{center}

Since for 
fusion plasmas the thermal conduction along $\bf B$ is expected
to  be fast in relation to  the heat transport perpendicular
to a magnetic surface,  equilibria
with  isothermal magnetic surfaces are a reasonable approximation
\cite{MaPe,ThPa,ClFa,Ta96,ThTa,TaTh}.
In particular, the even simpler case of 
isothermal resistive equilibria 
has also been considered \cite{GrHo}.
 
For $T=T(\psi)$ integration of Eq. (\ref{13}) leads to 
\begin{equation}
\frac{K^2B^2}{2\rho^2} + \lambda T \ln \rho =H(\psi),
						 \label{20}
\end{equation}
where $\lambda$ is the proportionality constant in the 
ideal gas law $P= \lambda \rho T$.
Consequently, Eq. (\ref{14}) reduces to
\begin{equation}
\nabla\cdot \left\lbrack \left(1 - \frac{K^2}{\rho}\right)
      \frac{\nabla \psi}{R^2} \right\rbrack
 + ({\bf v}\cdot {\bf B}) K^\prime + \frac{B_\phi}{R} X^\prime
 + \rho H^\prime -\lambda\rho(1 -\log \rho) T^\prime =0.  
						      \label{21}
\end{equation}
We remark that apart from the fact that the S terms have been replaced
by T terms, Eqs. (\ref{20}) and (\ref{21}) are identical with the
respective Eqs. (\ref{18}) and (\ref{19}).

\begin{center}
{\large\em C.\ \ Incompressible flows}
\end{center}

The existence of hyperbolic regimes
may be dangerous for plasma confinement because they are associated with 
shock waves which can cause equilibrium degradation. 
In this respect incompressible flows are of particular interest because,
as is well known from gas dynamics, it is the compressibility
that can give rise to shock waves; thus for incompressible flows
the equilibrium equation becomes always elliptic. 
For $\nabla \cdot {\bf v}=0$ it follows from
Eqs. (\ref{1}) and (\ref{9}) that the density is a surface quantity
\beq
\rho=\rho(\psi),
							\label{22}
\eeq
consistent with the fact that in fusion experiments
equilibrium density gradients parallel to $\bf B$ have not been
observed.

 With the aid of Eq. (\ref{22}),  integration of Eq. (\ref{13})
 yields an expression for the pressure:
\begin{equation} 
P = P_s(\psi) - \frac{v^2}{2} = P_s- \frac{K^2B^2}{2\rho}. 
						\label{23}    
\end{equation}
We note here that, unlike in static 
equilibria, in the presence of flow  magnetic surfaces in general
do not coincide with
isobaric surfaces  because Eq. (\ref{2})
implies that ${\bf B} \cdot {\bf \nabla} P$ in
general differs from zero.
In this respect, the term $P_s(\psi)$ is the static part of 
the pressure which
does not vanish  when ${\bf v} = {\bf 0}$.  
If it is now assumed that
$\frac{\textstyle K^2}{\textstyle \rho}\neq 1$
and Eq. (\ref{23}) is inserted into Eq. (\ref{14}), 
the latter reduces to
the {\em elliptic} differential equation
\begin{equation} 
(1-M^2) \Delta^\star \psi - 
	     \frac{1}{2}(M^2)^\prime |\nabla \psi|^2 
+ \frac{1}{2}\left(\frac{X^2}{1-M^2}\right)^\prime
+ R^2P_s^\prime   = 0.
						    \label{26}
\end{equation}
Eq. (\ref{26}) is identical in form to the corresponding
ideal equilibrium equation (Eq. (22) of Ref. \cite{TaTh}).
It is also noted that special cases of incompressible ideal
equilibria have been investigated in Refs. \cite{AvBh} and 
\cite{AnChe}.
Unlike to the corresponding sets
of compressible $S=S(\psi)$ equations  (\ref{18}) and (\ref{19}),  
and $T=T(\psi)$ equations (\ref{20}) and (\ref{21}),
Eq. (\ref{26}) is decoupled from Eq. (\ref{23}).
Once the solutions of Eq. (\ref{26}) are known, 
Eq. (\ref{23}) only determines the pressure. 

\begin{center}
{\large\bf III.\ \ The existence of solutions in relation to the  
		     conductivity profile}
\end{center}

We shall show that  the  compatibility of Eq. (\ref{12}) containing 
the conductivity $\sigma$ with the ``ideal" equations (\ref{13}) and (\ref{14}) 
depends crucially on the spatial
dependence of $\sigma$. 
In this respect the cases $\sigma=\sigma(R, \psi$),
and $\sigma=\sigma(\psi)$ are  examined below.

\begin{center}
{\large\em A.\   $\sigma=\sigma(R,\psi)$}
\end{center}

An explicit spatial dependence of  $\sigma$, in addition 
to that of $\psi$, is interesting because
it makes the equilibrium problem 
well posed, i.e. 
in this case Eq. (\ref{12}) can be 
 decoupled from the other Eqs. (\ref{13})
and (\ref{14}). A possible explicit spatial dependence
of $\sigma$  can be justified by the following arguments: 
(a) Even in Spitzer conductivity, 
$\sigma= \alpha T_e^{3/2}$, the quantity $\alpha$ has a (weak)
spatial dependence  and (b) 
cylindrically  symmetric resistive $\sigma=\sigma(\psi)$ equilibria
are possible \cite{Th}  and therefore the non-existence of   
axisymmetric static toroidal $\sigma=\sigma(\psi)$  equilibria
is related to the toroidicity involving through
the scale factor $|\nabla \phi|=1/R$;  this could also imply an 
explicit dependence of $\sigma$ on $R$.
In addition,  we may remark that the neoclassical conductivity 
depends on the aspect ratio $\cal A$ because
the fraction of trapped particles relates to $\cal A$ 
(see \cite{SaAn} and Refs. cited
therein). 
It should be noted, however, that
 a knowledge of the  $\sigma $-profile in the various 
collisionality regimes  of magnetic confinement has not been obtained
to date. 
 
For us the main advantage in allowing 
    $\sigma=\sigma(R,\psi)$ lies in the fact that
 Eq. (\ref{12})                       
 can then be considered as a formula 
 determining the conductivity
\beq 
\sigma = \frac{\Delta^\star \psi}{V_c}, 
						   \label{26a}
\eeq
provided  $\psi$ is known. Also, the poloidal electric field can then  be 
obtained by Eq. (\ref{11}).

To determine $\psi$ in the case of compressible  flows with isentropic magnetic 
surfaces the set of Eqs. (\ref{18}) and (\ref{19}),
which are coupled through the density
$\rho$, should be solved numerically under appropriate boundary conditions.
This can be accomplished by the existing ideal MHD 
equilibrium codes \cite{SeGr,KeTo,ZeSt}. 
The problem of compressible flows with isothermal magnetic
surfaces [Eqs. (\ref{20}) and (\ref{21})] can be solved in a similar way.

For incompressible flows $\psi$ can be determined by Eq. (\ref{26}) alone,
which is amendable to several classes of analytic solutions.
In particular, sheared- poloidal-flow  equilibria associated
with ``radial" (poloidal) electric fields which play a role 
in the L-H transition can be constructed by means of the transformation
\cite{Cl,Mo}
\beq
U(\psi)= \int_0^{\psi}\, [1-M^2(\psi^{\prime})^{1/2}]\, d\psi,\ \    M^2<1,
						       \label{27}
\eeq
Under this transformation Eq. (\ref{26})   
reduces (after dividing by $(1-M^2)^{1/2}$) to 
\begin{equation} 
 \Delta^\star U  
 + \frac{1}{2}\frac{d}{d U} \left(\frac{X^2}{1-M^2}\right)
+   R^2\frac{dP_s}{d U}  =0.
							 \label{28}
\end{equation}
It is noted here that the requirement $M^2<1$ in transformation (\ref{27})
implies that $v_{p}^2<v_s^2$, where $v_s= (\gamma P/\rho)^{1/2}$ is the
sound speed. This follows from 
Eqs. (\ref{25}) and (in Gaussian units) 
$$
\left(\frac{v_s}{v_{Ap}}\right)^2 = 
(\gamma/2)\frac{8\pi P}{h^2|\nabla \psi|^2}\approx 1.
							 \label{29}
$$ 
Since, 
according to experimental evidence in tokamaks \cite{Bu},  
 the (maximum) value
of the ion poloidal velocity in the edge region 
during the L-H transition
is of the order of 10 Km/sec and the ion temperature is of the order 
of $1$ KeV, the scaling $v_p \ll v_s$ is satisfied in this region.
Therefore, 
the restriction $M^2<1$ is of non-operational relevance.  
The simplest solution of Eq. (\ref{26}) corresponding to $M^2=$ const., $X^2=$ const. 
and $P_s\propto\psi$, is  given by
\beq
\psi=\psi_c \left(\frac{R}{R_c}\right)^2
     \left\lbrack2-\left(\frac{R}{R_c}\right)^2 
     - d^2\left(\frac{z}{R_c}\right)^2 \right\rbrack,       \label{30}
\eeq
where $\psi_c$ is the $\psi$ value on the magnetic axis located
at ($z=0$, $R=R_c$) and $d$ is a parameter related to the 
shape of flux surfaces.
Equation (\ref{30}) describes the Hill's vortex configuration \cite{Tho}.
The conductivity then follows from Eq. (\ref{26a}):
\beq
\sigma=\sigma_c \left(\frac{R}{R_c}\right)^4
     \left\lbrack2-\left(\frac{R}{R_c}\right)^2 
     - d^2\left(\frac{z}{R_c}\right)^2 \right\rbrack,
							   \label{31}
\eeq
where $\sigma_c$ is the value of $\sigma$ on the magnetic axis.
The conductivity profile in the middle-plane $z=0$ is illustrated
in Fig. 1. We remark the  outward displacement of the
maximum-conductivity position $R_{\max}$  with respect to $R_c$
($R_{\max}/R_c = 2/\sqrt 3$)
and the  asymetry of the inner part of the profile
as compared with  the outer part due to the explicit $R$ dependence 
of $\sigma$.

\begin{center}
{\large\em B. \ \ $\sigma=\sigma(\psi)$}
\end{center}
For this case we consider Eq. (\ref{12})  in the vicinity 
of the magnetic axis by transforming the coordinates from ($R, z, \phi$)
to ($x,y, \phi$) (Fig. 2).
The transformation is given by
\begin{eqnarray}
R&=&R_c+x= R_c + r cos\theta  \nonu
z&=&y=-rsin\theta.
			      \label{32}
\end{eqnarray}
The quantities $\psi(x,y)$ and  $\sigma(\psi)$  are then 
expanded to second-order in $x$ and $y$: 
\beq
\psi(x,\psi) = \psi_c + c_1\frac{x^2}{2} + c_2\frac{y^2}{2}
	       + c_3 xy + \ldots
						 \label{33}
\eeq
and 
\beq
\sigma = \sigma_c + \sigma_1 (\psi-\psi_c) + \ldots
       = \sigma_c + \sigma_1 (c_1\frac{x^2}{2} + c_2\frac{y^2}{2}
	       + c_3 xy + \ldots) + \ldots.
							\label{34}
\eeq
Here, $c_1=(\pars^2 \psi/\pars x^2)_c $, $c_2=(\pars^2 \psi/\pars y^2)_c$,  
$c_3=(\pars^2 \psi/\pars x\pars y)_c $, $\sigma_c$ is the conductivity
on the magnetic axis and $\sigma_1=$ const.
On the basis of Eqs. (\ref{33})  and (\ref{34}) Eq. 
$\Delta^\star\psi=V_c \sigma(\psi)$
becomes a polynomial in $x$ and $y$ which should vanish identically.
This requirement leads to $c_1=c_3=0$ and, therefore, it follows from
Eq. (34) that the magnetic surfaces in the vicinity of the 
magnetic axis are not closed surfaces. 

The non-existence of $\sigma(\psi)$ equilibria with closed magnetic
surfaces can be extended to the case of non-parallel flows lying
within the magnetic surfaces. Indeed, if the relation 
${\bf v}\cdot \nabla \psi=0$
is assumed instead of ${\bf v}\paral {\bf B}$, the toroidal 
component of Eq. (\ref{6}) leads again to Eq. (\ref{12}).

A possible proof of the non-existence of $\eta=\eta(\psi)$ equilibria 
far from the magnetic axis has not been obtained to date. It may
be noted, however, that for  $\sigma=\sigma(\psi)$, 
Eq. (\ref{14}) becomes 
{\em parabolic}. This follows by considering in this equation
the  determinant
$\cal D$ of the symmetric matrix of coefficients. 
On account  of 
$\Delta^\star\psi = V_c \sigma(\psi)$, and 
$\rho= \rho (R, \psi, |\nabla\psi|)$
by Eq.  (\ref{13}), the second derivatives   
of equation (\ref{14}) are contained only in the term 
$$
\frac{K^2}{\rho}\frac{\pars \rho}{\pars |\nabla \psi|^2}
\nabla|\nabla \psi|^2 \cdot \nabla \psi,
$$
which comes from the term  $\nabla\cdot[(1-K^2/\rho)\nabla \psi/R^2]$.
Subsequent  evaluation  of  $\cal D$  leads to  ${\cal D }=0$.
Therefore, the function $\psi$ is (over)restricted everywhere 
to satisfy a parabolic equation and the elliptic 
equation $\Delta^\star\psi = V_c \sigma(\psi)$. 

\begin{center}
{\large\bf IV.\ \ Conclusions}
\end{center}
The equilibrium of an axisymmetric plasma with flow parallel
to the magnetic field has been investigated 
within the framework of the resistive magnetohydrodynamic (MHD) theory.
For the system under consideration the equilibrium equations
reduce to a set of a second-order differential equation for the
poloidal magnetic flux function $\psi$ coupled through the density
with an algebraic Bernoulli
equation, which  are identical in form with the corresponding 
ideal MHD equations, and the equation $\Delta^\star\psi= V_c \sigma$.
($\Delta^\star$, $V_c$ and $\sigma$  are the Grad-Schl\"uter-Shafranov 
elliptic operator,  the constant toroidal loop voltage and 
the conductivity, respectively. The existence of solutions of the above
mentioned set of equations is sensitive to the spatial dependence 
of $\sigma$. 

For a conductivity of the form $\sigma=\sigma(R, \psi)$,   
Eq. $\Delta^\star\psi= V_c \sigma$ can be considered  
uncoupled to the other two equations, thus determining only
the conductivity. 
For  compressible flows  
and isentopic magnetic surfaces 
 the differential
equation for $\psi$ [(Eq. (\ref{19})], 
pending on the value  of the poloidal flow,   
can be either elliptic or hyperbolic. 
Solutions of the set of this equation and the
coupled  Bernoulli equation [Eq. (\ref{18})] can be 
 obtained numerically. 
The problem of compressible equilibria
with isothermal magnetic surfaces [Eqs. (\ref{20}) 
and (\ref{21})] can be solved in a similar way. 
For incompressible equilibria $\psi$ obeys an elliptic differential
equation [(Eq. (\ref{26})],  uncoupled to the associated Bernoulli equation 
[Eq. (\ref{23})] which just determines the pressure. 
 Several classes of analytic equilibria with incompressible flows 
 having qualitatively plausible $\sigma$ profiles,
i.e, profiles with  
 $\sigma$ taking a  maximum value  close to the magnetic axis and a minimum 
 value on the plasma surface, can be constructed. In particular, 
 sheared-poloidal-flow 
equilibria can be derived  by means of the  transformation (\ref{27})
for $\psi$.

For $\sigma=\sigma(\psi)$ appreciation of $\Delta^\star\psi= V_c \sigma$
in the vicinity  of the magnetic axis proves 
therein, irrespective of plasma  compressibility,  
the non-existence of closed magnetic surfaces. 
This result can be extended to the case of non-parallel flows lying within  
the magnetic surfaces. In addition,  for parallel flows $\psi$
is (over)restricted to satisfy  throughout the plasma
an elliptic 
and a parabolic differential equations.

According to the results of the present investigation, 
the existence of resistive equilibria is sensitive to the spatial 
dependence of conductivity. 
Thus, the task of obtaining this dependence in the various 
confinement regimes of fusion plasmas may deserve 
further experimental 
and theoretical investigations. A conductivity with a 
spatial dependence in addition to that of $\psi$, on the one hand, 
would open up the possibility of the  existence of several classes 
of resistive equilibria free of Pfirsch-Schl\"uter diffusion. On the other 
hand, a strict Spitzer-like conductivity, $\sigma=\sigma(\psi)$, should 
imply the persistence of a Pfirsch-Schl\"uter-like diffusion also in 
the non-linear flow regime. 

\begin{center}
 {\large\bf Acknowledgments}
\end{center}

Part of this work was conducted during a visit by one of the authors 
(G.N.T.) to  the Max-Planck Institut  f\"ur Plasmaphysik, Garching.
The hospitality of that Institute is greatly appreciated.

\newpage
\begin{center}
{ \large \bf Figure captions}
\end{center}

\begin{description}
\item[FIG. 1.]
	The conductivity profile on the middle-plane $z=0$ described by 
         Eq. (\ref{31})

\item[FIG. 2.] The system of coordinates ($x,y,\phi$).
\end{description}
\newpage
\thispagestyle{empty}
\begin{center}
\vspace{10cm}
\begin{figure}[htb]
\epsfbox{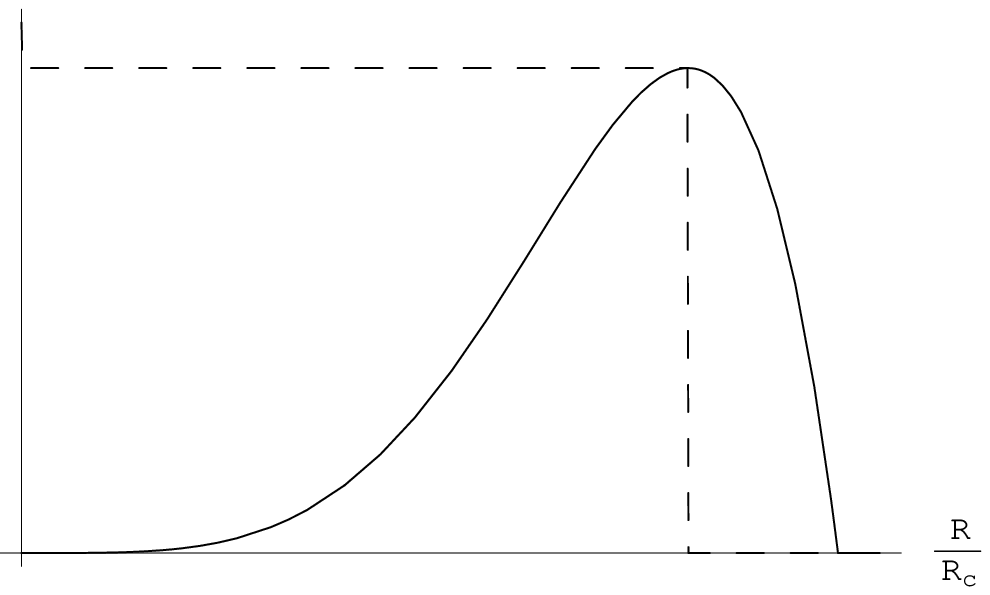}
\vspace{-12.9cm}

\hspace*{6.4cm}$2/\sqrt{3}$ \hspace{0.6cm} $\sqrt{2}$
\vspace*{-6.2cm}

\noindent
\hspace*{-1.2cm} 32/27

\vspace{-1.8cm}
{\Large $\frac{\sigma}{\sigma_c}$}
\end{figure}
\end{center}
\vspace{6.5cm}

\noindent
\hspace*{-1cm}FIG. 1. \        The conductivity profile on the middle-plane $z=0$ described by
         Eq. (\ref{31})
\newpage
\thispagestyle{empty}  
\begin{center}
\begin{figure}[htb]
\epsfbox{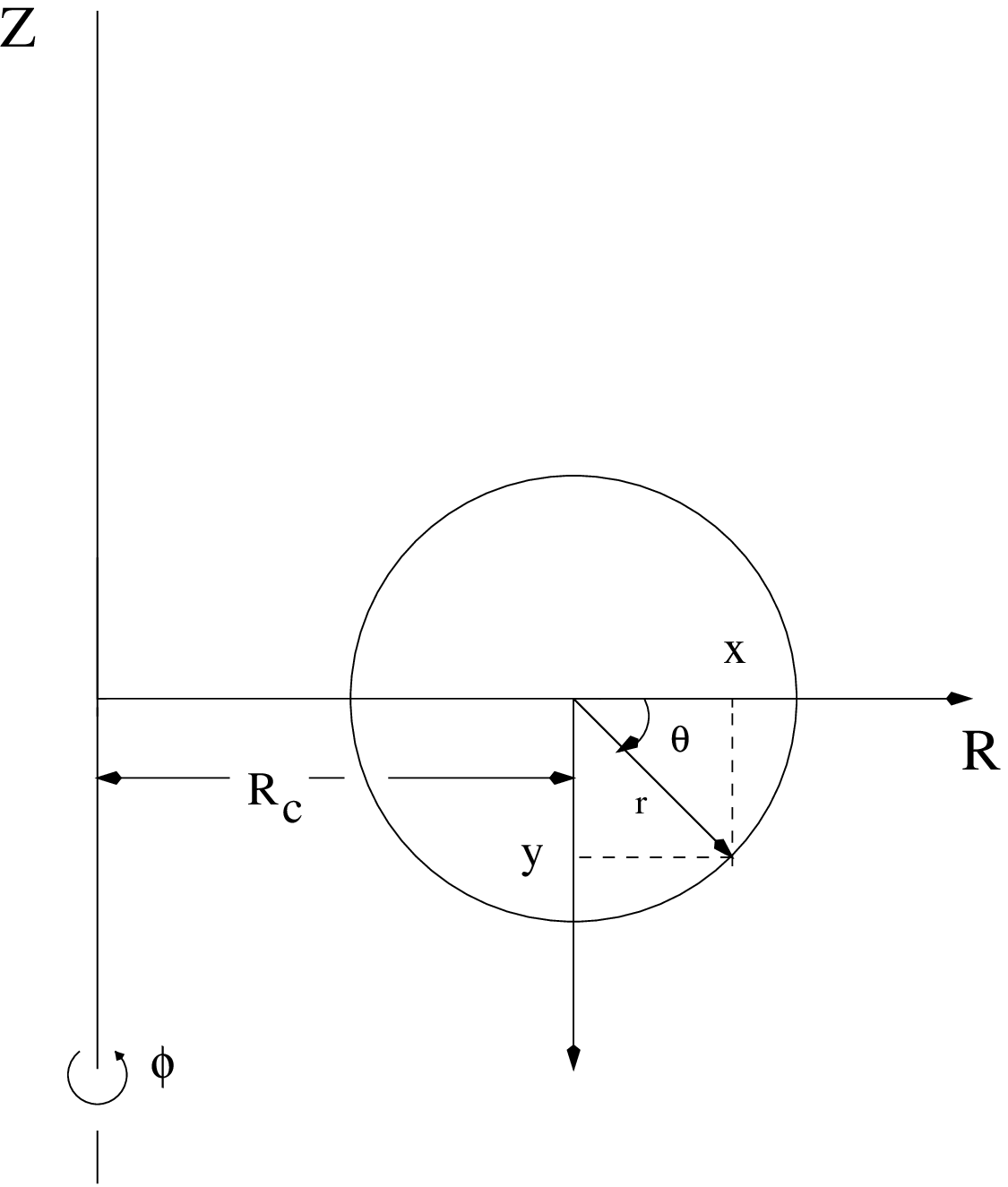}
\end{figure}
\end{center}
\noindent
FIG. 2. \  The system of coordinates ($x,y,\phi$).

\begin{thebibliography}{99}
\bibitem{MoVo} D. Moreau    and I. Voitsekhovitch, Nucl. Fusion {\bf 39},
	       685 (1999).
\bibitem{Ta} H. Tasso, {\it Lectures on Plasma Physics},
	     Report  IFUSP/P-181, LFP-8,  Universidade 
	     de S\~ao Paulo, Instituto de F\'isica, S\~ao Paulo (1979).
\bibitem{MoSh} D. Montgomery, and X. Shan, Comments Plasma Phys. Contolled 
		 Fusion {\bf 15}, 315 (1994).
\bibitem{BaLe} J. W. Bates and H. R. Lewis, Phys. Plasmas {\bf 3}
	       2395 (1996).
\bibitem{MoBh} D. Montgomery, J. W. Bates, and H. R. Lewis,  
		Phys. Plasmas {\bf 4}, 1080 (1997).
\bibitem{SuEu}  S. Suckewer, H. P. Eubank, G. J. Goldston 
		E. Hinnov and N. R. Sauthoff,   Phys. Rev. Lett. {\bf 43}, 207 (1979).
\bibitem{BrBi}  K. Brau, M. Bitter, R. J. Goldston, D. Manos K. McGuire, 
		S. Suckewer, Nucl. Fusion {\bf 23}, 1643 (1983).
\bibitem{TaDo}  H. F. Tammen, A. J. H. Donn\'e, H. Euringer and T. Oyevaar,   
	       Phys. Rev. Lett. {\bf 72}, 356 (1994).
\bibitem{Th}   G. N. Throumoulopoulos, J. Plasma Physics {\bf 59},
	       303 (1998).
\bibitem{ZeGr} H. P. Zehrfeld and B. J. Green, Nucl. Fusion {\bf 12}, 
	       569 (1972).  
\bibitem{MoSo} A. I. Morozov and L. S. Solov\'{e}v, {\it Reviews of 
	      Plasma Physics}  {\bf  8}, 1 (1980), edited by M. A. Leontovich
	      (Consultants Bureau, New York).
\bibitem{Ha}  E. Hameiri, Phys. Fluids {\bf 26}, 230 (1983). 
\bibitem{SeGr} S. Semenzato, R. Gruber and H. P. Zehrfeld, Comput. Phys. 
	       Rep. {\bf 1}, 389 (1984).
\bibitem{KeTo}  W. Kerner,  and S. Tokuda,   Z. Naturforsch. {\bf 42a},
	       1154 (1987) 
\bibitem{ZeSt} R. \.Zelazny, R. Stankiewicz, A. Galkowski 
	       and S. Potempski  {\it et al.},
		Plasma Phys. Contr. Fusion {\bf 35}, 1215 (1993).
\bibitem{Ja} J. D. Jackson {\it Classical Electrodynamics}, Second Edition
	       (John Wiley \& Sons, New York, 1975) p. 335.
\bibitem{MaPe} E. K. Maschke and H. Perrin, Plasma Phys. 
	       {\bf 22}, 579 (1980).
\bibitem{ThPa} G. N. Throumoulopoulos and G. Pantis, Phys. Plasmas B {\bf 1}, 
		1827 (1989).  
\bibitem{ClFa} R. A. Clemente and R. Farengo, Phys. Fluids {\bf 27}, 776
	       (1984).
\bibitem{Ta96} H. Tasso,  Phys. Lett. A {\bf 222},  97 (1996).  
\bibitem{ThTa} G. N. Throumoulopoulos and H. Tasso, Phys. Plasmas {\bf 4},  
	     1492 (1997).  
\bibitem{TaTh} H. Tasso and G. N. Throumoulopoulos, Phys. Plasmas {\bf 5},  
	     2378 (1998).  
\bibitem{GrHo}  H. Grad and J. Hogan,  Phys. Rev. Lett. {\bf 24}, 1337
	       (1970).
	     \bibitem{AvBh} K. Avinash, S. N. Bhattacharyya and B. J. Green,
	       Plasma Phys. Control. Fusion {\bf 34}, 465 (1992). 
\bibitem{AnChe} Zh. N. Andruschenko, O. K. Cheremnykh and 
		J. W. Edenstrasser,  J. Plasma Physics   {\bf 58}, 
		421 (1997).  
\bibitem{SaAn} O. Sauter, C. Angioni and Y. R. Lin-Liu, 
	       Phys. Plasmas {\bf 6}, 2834 (1999). 
\bibitem{Cl} R. A. Clemente, Nucl. Fusion {\bf 33}, 963 (1993).    
\bibitem{Mo} P. J. Morrison,  Private communication; transformation 
		   (\ref{27}) was discussed in the invited talk entitled 
		   ``A generalized energy principle" which was delivered
		  in the Plasma-Physics  APS Conference, Baltimore 1986.
\bibitem{Bu} K. H. Burrell, Phys. Plasmas {\bf 4}, 1499 (1997). 
\bibitem{Tho} W. B. Thompson, {\it An introduction to Plasma Physics}
	      (Addison-Wesley, Reading, Massachusetts, 1964), p. 55.   
%
%
%
\end{thebibliography}
\end{document}